\documentclass[aps,prc,twocolumn,floatfix,superscriptaddress]{revtex4}
\usepackage{epsfig}
\usepackage{amsmath}
\usepackage{color}

\usepackage{graphicx}

\begin{document}

\title{Longitudinal asymmetry in heavy ion collisions at RHIC} %in Cu+Cu and Au+Au collision at 200 GeV and 62.4 GeV}
\author{Sanchari Thakur}
\email{s.thakur@vecc.gov.in}
\affiliation{Variable Energy Cyclotron Centre, HBNI, 1/AF, Bidhan Nagar, Kolkata-700064, India}

\author{Sumit Kumar Saha}
\email{sk.saha@vecc.gov.in}
\affiliation{Variable Energy Cyclotron Centre, HBNI, 1/AF, Bidhan Nagar, Kolkata-700064, India}

\author{Sk Noor Alam}
\email{noor1989phyalam@gmail.com}
\affiliation{Department of Physics, Aligarh Muslim University, Aligarh-202002, India}

\author{Rupa Chatterjee}
\email{rupa@vecc.gov.in}
\affiliation{Variable Energy Cyclotron Centre, HBNI, 1/AF, Bidhan Nagar, Kolkata-700064, India}

\author{Subhasis Chattopadhyay}
\email{sub@vecc.gov.in}
\affiliation{Variable Energy Cyclotron Centre, HBNI, 1/AF, Bidhan Nagar, Kolkata-700064, India}

\begin{abstract}

The longitudinal asymmetry arises in relativistic heavy ion collisions due to fluctuation in the number of participating nucleons. This asymmetry  causes a  shift in the center of mass rapidity of the participant zone. The rapidity shift as well as the longitudinal asymmetry have been found to be significant at the top LHC energy for collisions of identical nuclei. We study the longitudinal asymmetry and its effect on charged particle rapidity distribution and anisotropic flow parameters at relatively lower RHIC energies using a model calculation. 
The rapidity shift is found to be more pronounced for peripheral collisions, smaller systems and also for lower beam energies due to longitudinal asymmetry. A detailed study has been done by associating the  average rapidity shift to a polynomial relation where the coefficients of this polynomial characterize the effect of the asymmetry. We show that the rapidity shift may affect observables significantly at RHIC energies.

\end{abstract}
%\pacs{25.75.-q,12.38.Mh}

\maketitle

\section{Introduction} 
In a heavy ion  collision apart from  participants there are  spectator nucleons  which simply pass through the collision zone undeflected without any interaction~\cite{qgp1, qgp2, qgp3, qgp4}.  The number of participating nucleons from each nucleus is expected to be different in a collision of non-identical nuclei. However, in a collision of identical nuclei  also the number of participants fluctuates event-by-event  due to fluctuations in the positions of the nucleons around the mean nuclear density profile. This inequality in the  participant number results in  a non-zero net momentum of the nucleon-nucleon centre of mass frame. As a result, the  center of mass of participants does not coincide with the collider center of mass of the system. This also causes a  rapidity shift of the participant zone with respect to the nucleon-nucleon centre of mass rapidity \cite{Vovchenko}.  This is known as longitudinal asymmetry where the magnitude of rapidity shift $y_{0}$ characterizes the longitudinal asymmetry~\cite{raniwala, Vovchenko, alice, r_raniwala}.
If A and B are the number of  participants from two nuclei, the longitudinal asymmetry in participants is defined as, $\alpha_{part} = (A-B)/(A+B)$. The value of $\alpha_{part}$ can be significantly large for collisions of non-identical nuclei. The rapidity shift caused by the net momentum  of the participant zone  can be approximated as~\cite{alice, raniwala},
\begin{eqnarray}
 y_{0} \approx \frac{1}{2} ln \frac{A}{B} \ .
\end{eqnarray}
 The equation above is obtained considering  $m_{0} \ \ll \ p$ at RHIC ($m_0/p \ < $ 10$^{-4}$). The rapidity shift can be expressed in terms of the longitudinal asymmetry as,
 \begin{eqnarray}
 y_{0} = \frac{1}{2} ln \frac{1+\alpha_{part}}{1-\alpha_{part}} .
 \end{eqnarray}
 
For small $\alpha_{part}$, $y_{0} \approx \alpha_{part}$. 

The longitudinal asymmetry in spectator is defined as,
\begin{eqnarray}
 \alpha_{spec} &=& \frac{(N-A) - (N-B)}{(N-A) + (N-B)} = \frac{B-A}{2N-(A+B)} \\
               &=& - \alpha_{part} \frac{A+B}{2N-(A+B)} 
\end{eqnarray}
where N is the total number of nucleons in each nuclei and (N-A) and (N-B) are the number of spectators.

%The spectator asymmetry and  participant asymmetry are related as 
%\begin{equation}
% \alpha_{spec} = - \alpha_{part} \frac{A+B}{2N-(A+B)} ,
%\end{equation}
The rapidity shift is related to spectator asymmetry as, 
\begin{eqnarray}
y_{0} = \frac{1}{2} ln \frac{(A+B)(1+\alpha_{spec})-2N\alpha_{spec}}{(A+B)(1-\alpha_{spec})+2N\alpha_{spec}} .
\end{eqnarray}

We know that the number of participants in a  heavy ion collision is not an experimentally measurable quantity. However, the fluctuation in the number of participants causes a fluctuation in the number of spectators. This can be probed by measuring the asymmetry in the energy deposition by the spectator nucleons in the two Zero-Degree Calorimeters (ZDCs) as
$\frac{E_{ZNA}-E_{ZNC}}{E_{ZNA}+E_{ZNC}}$~\cite{alice}. Therefore  the rapidity shift can be measured experimentally  by the spectator asymmetry rather than the participant asymmetry.

The longitudinal asymmetry is expected to affect the observables more
which depend directly on the rapidity, like the charged particle rapidity distribution. The  $dN_{\rm ch}/d\eta$ distribution is asymmetric in collisions such as d+Au or p+Pb and this has been described by the rapidity shift of the participant zone \cite{Steinberg,Garcia}.

However, the effect of longitudinal asymmetry on the  charged particle rapidity distribution has been found to be significant even for collisions of identical nuclei. This effect has been studied in Pb+Pb collisions at 2.76 A TeV  from the measurement of spectator energy deposition at the ZDCs by ALICE Collaboration~\cite{alice}  and also using model frameworks~\cite{raniwala}.  The  analysis by ALICE Collaboration shows that the longitudinal asymmetry  provides valuable information about the initial state and can be used to classify events~\cite{alice}.

The fluctuations in the fireball rapidity density is expected to produce nontrivial rapidity correlations. It was shown in Ref.~\cite{Bzdak} that the number of participants moving to left and right averaged over many events are equal and the rapidity distribution is symmetric (about y=0 in a symmetric collisions). However, this number is not same in a single event resulting in an asymmetric rapidity distribution. The rapidity correlation is shown to depend  on the rapidity difference and as well as on the  rapidity sum~\cite{Bzdak, krishnan}. The effect of longitudinal fluctuations on the azimuthal anisotropy coefficients and their rapidity dependence have been studied using a fluid dynamical simulation framework in Ref.~\cite{fluid}. The initial state longitudinal asymmetry has also been studied at FAIR energy recently~\cite{nbu}.

The effect of initial state fluctuations is expected to be more pronounced for peripheral collisions and also for smaller systems~\cite{rc}. Thus, the longitudinal asymmetry and rapidity shifts are also expected to be larger in smaller systems and at lower collision energies.

In the present work, we have used AMPT model~\cite{ampt}  (with string melting (SM)) to investigate  the effect of the longitudinal asymmetry  on the charged particle rapidity distribution and anisotropic flow parameters from Cu+Cu and Au+Au collisions at $\sqrt{s_{NN}}$= 200 GeV and 62.4 GeV at RHIC. 

The paper is organized as follows. In Section II, we have discussed the event by event rapidity shift, the spectator asymmetry distributions and also the event-by-event distributions of the rapidity shift with participant and spectator asymmetry. The effect of the shift on the particle rapidity distribution has been discussed in Section III using AMPT-SM for Cu+Cu and Au+Au at different centrality bins and at two different energies. In Section IV, the effect of the rapidity shift on elliptic  and triangular flow parameters has been shown and the results are summarized in Section V.

\section{Longitudinal asymmetry and rapidity shift}
\begin{figure}[h]
		\includegraphics[width=0.5\textwidth]{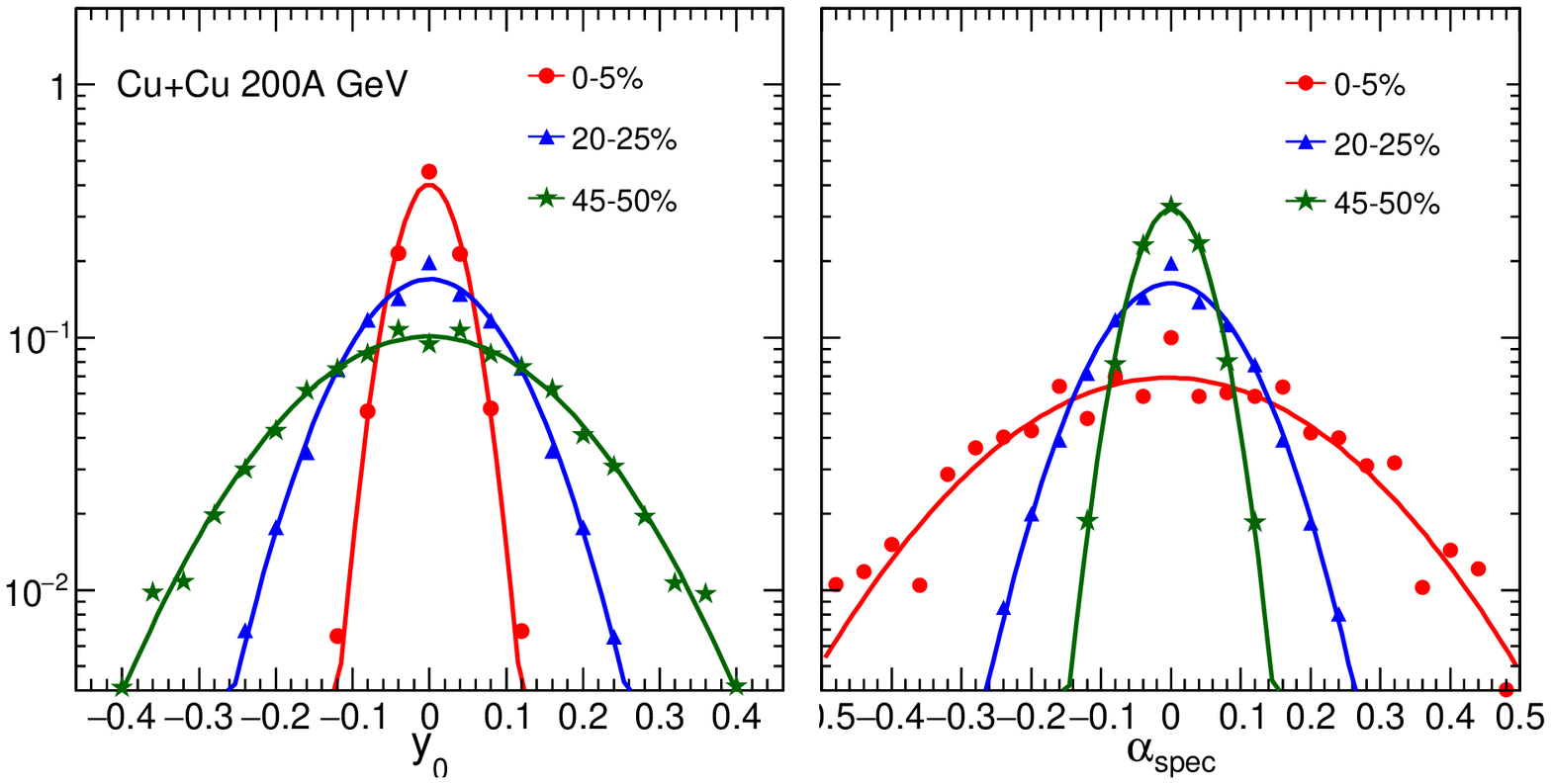}
		\hfill
			\includegraphics[width=0.5\textwidth]{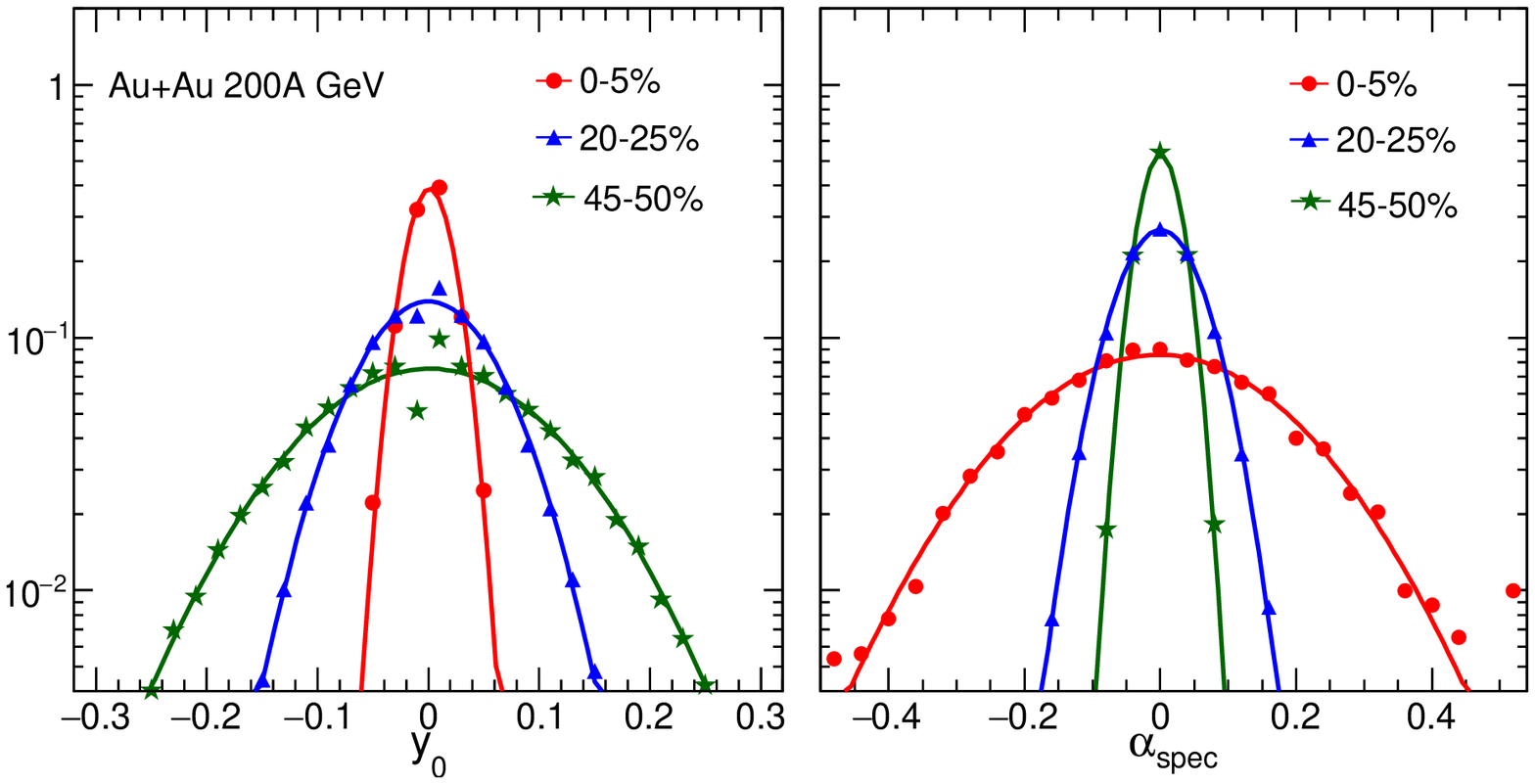}
		\caption{The distributions of the participant-zone rapidity shift $y_{0}$ and the spectator asymmetry $\alpha_{spec}$ for 0--5\%, 20--25\%, and 45--50\%  centrality classes of Cu+Cu [upper panels] and  Au+Au [lower panels] collisions at 200A GeV at RHIC. The y-axis represents the number of events (normalized).}
	\end{figure}

The distributions of $y_{0}$ and $\alpha_{spec}$ calculated using a Monte Carlo Glauber (MCG) initial state~\cite{glauber1,glauber2, glauber3, glauber4} are shown in Fig.1 for centrality bins 0--5\%, 20--25\% and 45--50\% along with  Gaussian fits  from Au+Au and Cu+Cu collisions at 200A GeV.

The width of the $y_{0}$ distribution increases from central to peripheral collisions. The relative fluctuation increases for larger impact parameters as the number of participants is relatively smaller for those collisions.

The events are divided into negative ($y_{0} <$ 0) and  positive ($y_{0} >$ 0) asymmetry events. The $\alpha_{part}$ distributions calculated using MC-Glauber are nearly identical to the $y_{0}$ distributions (as $y_{0} \approx \alpha_{part}$) however, the $\alpha_{spec}$ distributions are different.

An opposite trend between $y_{0}$ and $\alpha_{spec}$ distributions is observed.  The width of the $y_{0}$  distribution increases for peripheral collisions whereas, it decreases for $\alpha_{spec}$.  The relative fluctuation  decreases towards central collisions as the number of spectators decreases. In addition, the width of the distribution is found to be more for Cu+Cu collisions than for Au+Au collisions for a particular centrality bin.  It  indicates that the relative fluctuation increases with  decrease in system size as expected.

\begin{figure}[h!]
	\includegraphics[width=0.5\textwidth]{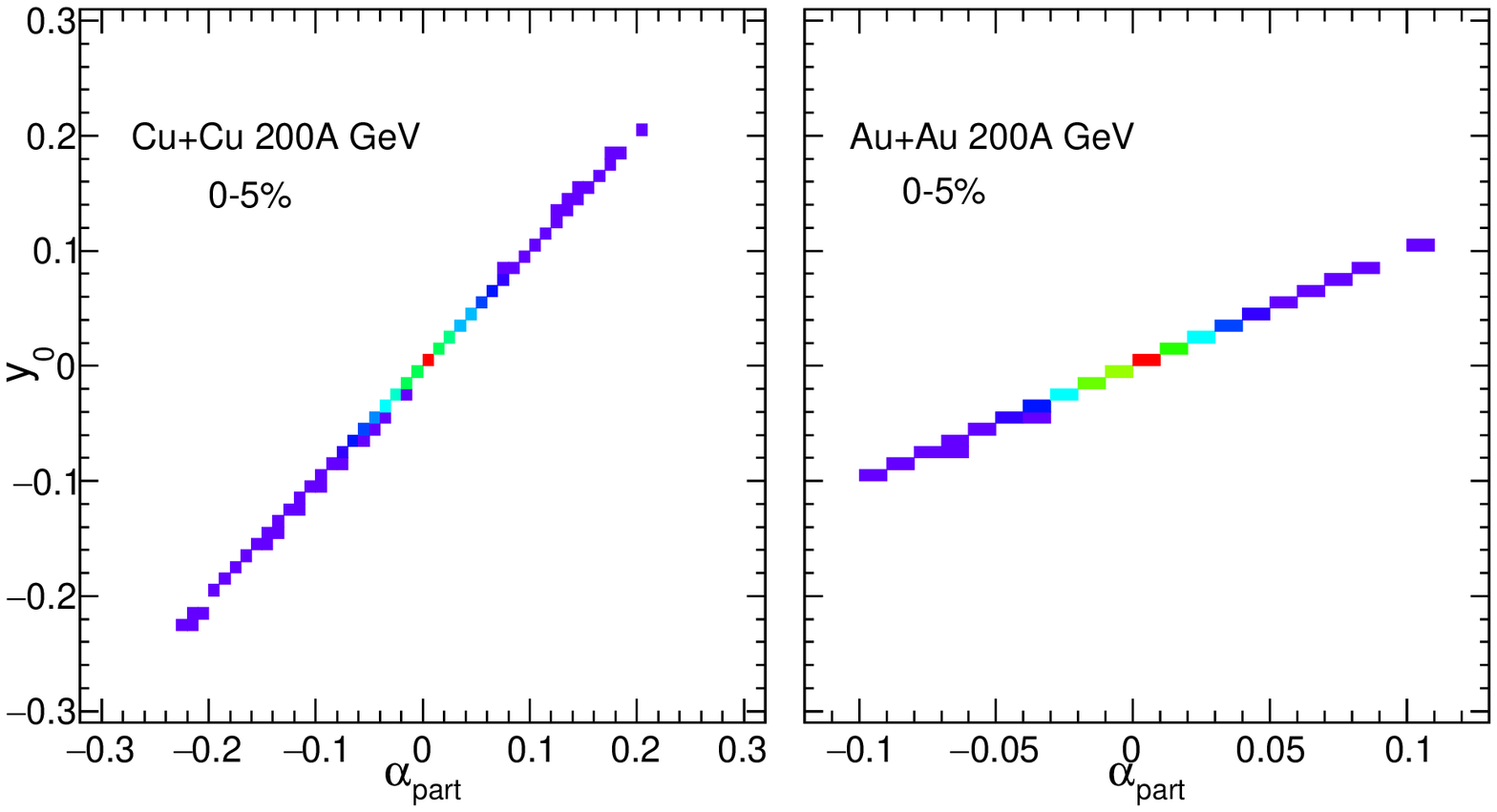}
	\vfill
	\includegraphics[width=0.5\textwidth]{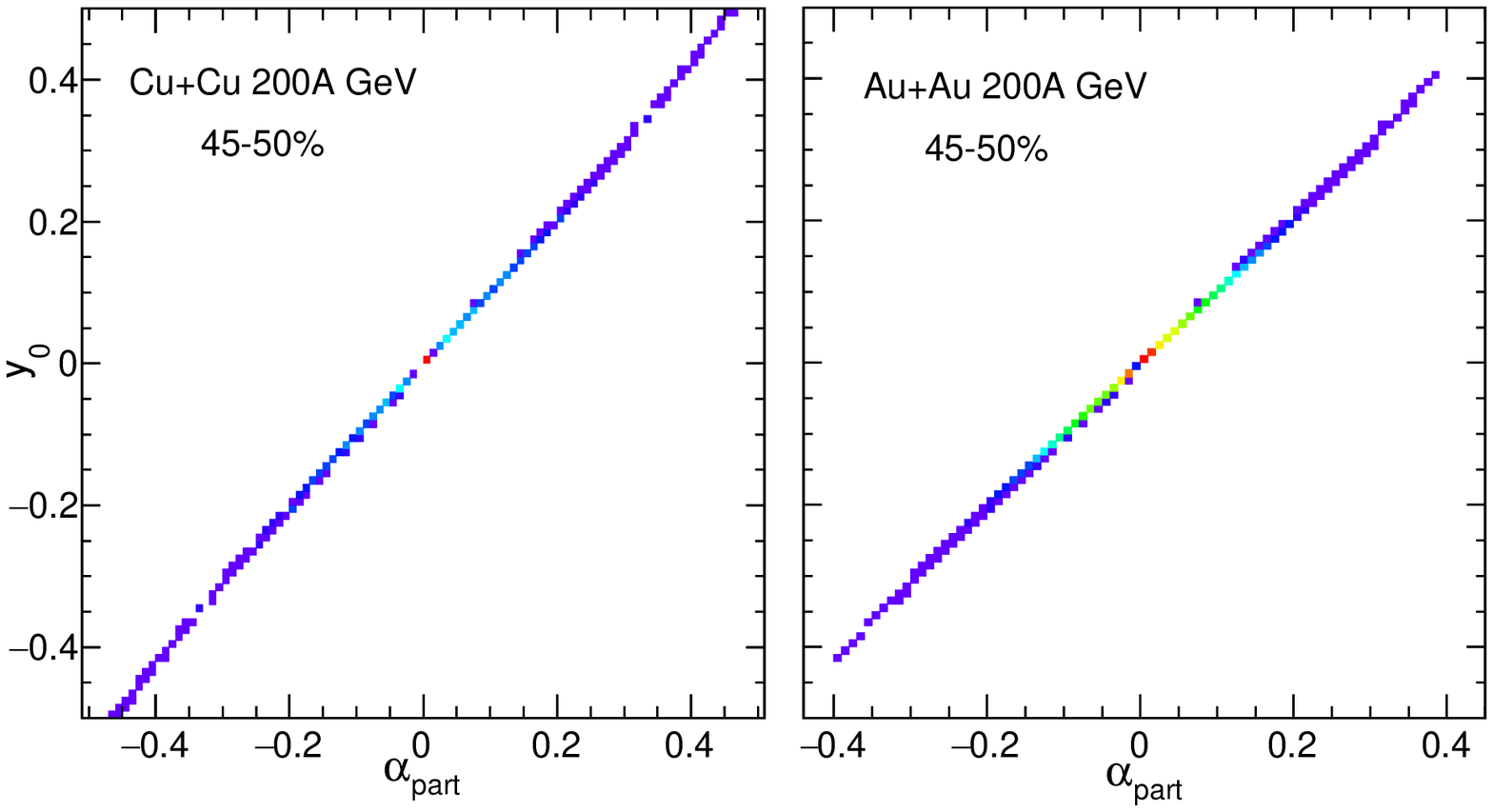}	
	\caption{Event-by-event distribution of $y_{0}$ with $\alpha_{part}$  from Cu+Cu and Au+Au at 200A GeV for 0--5\% (top) and 45--50\% (bottom) centralities.}
\end{figure}

\begin{figure}[h]
	\includegraphics[width=0.5\textwidth]{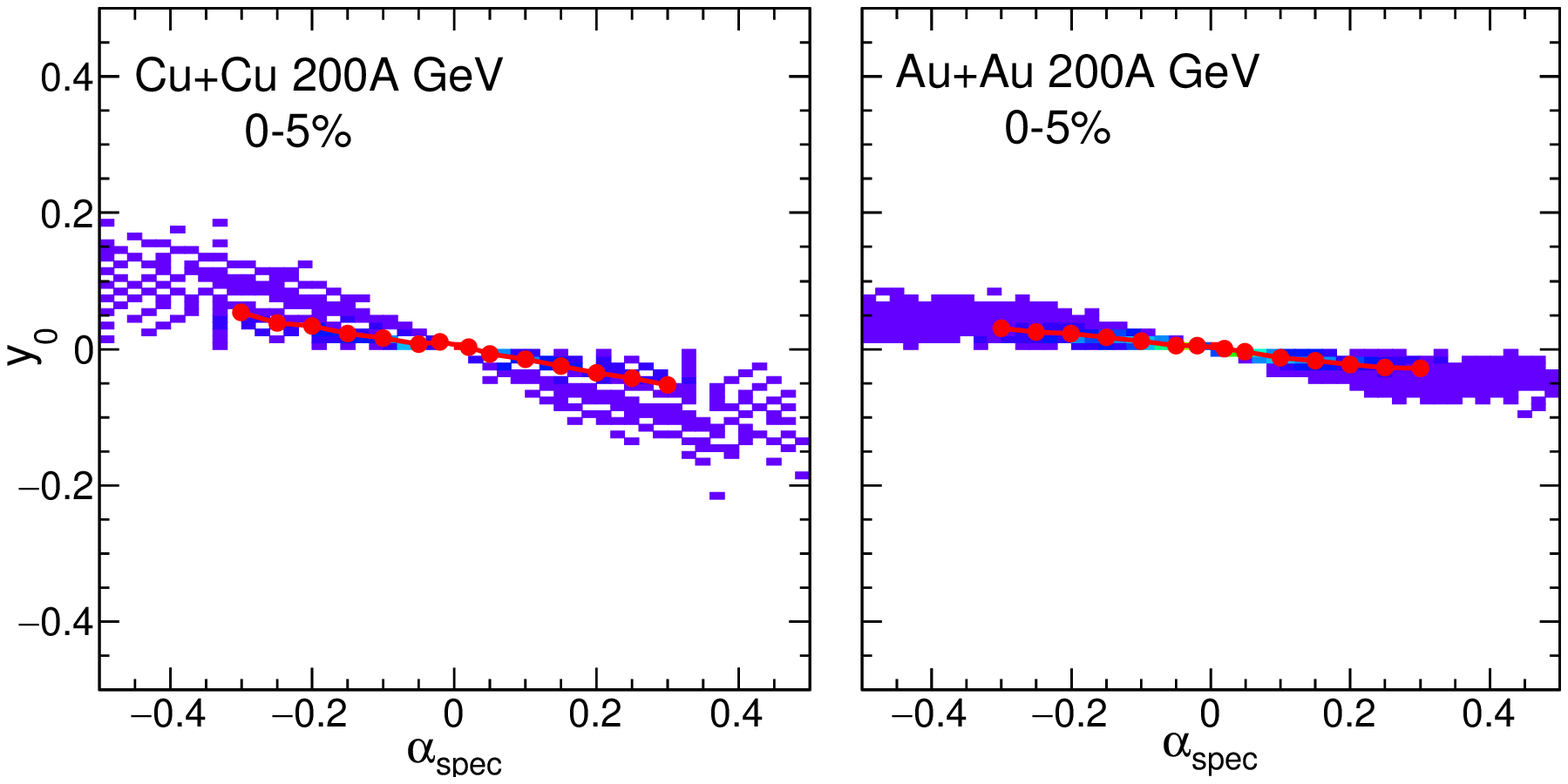}
	\vfill
	\includegraphics[width=0.5\textwidth]{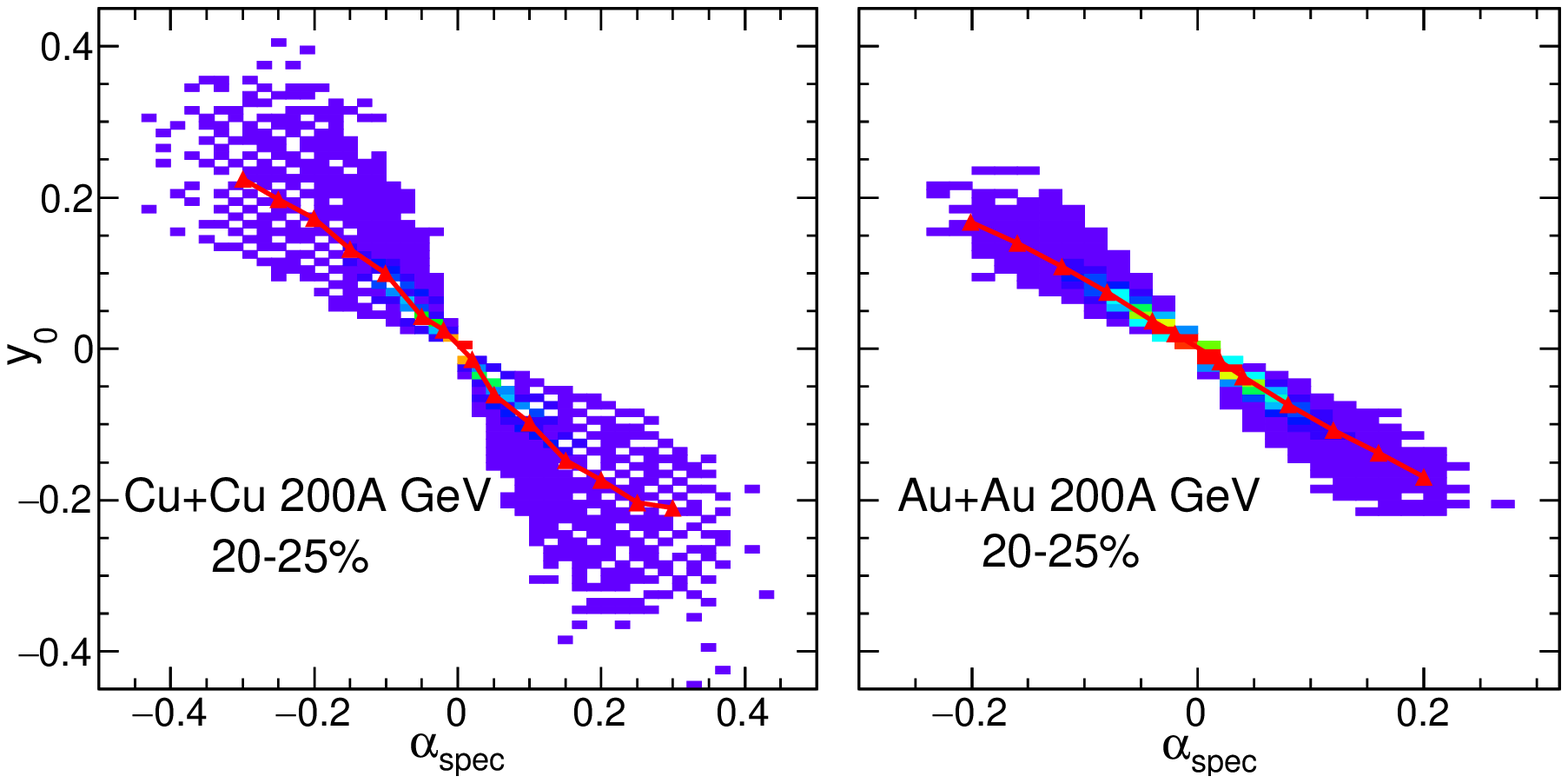}
	\vfill
	\includegraphics[width=0.5\textwidth]{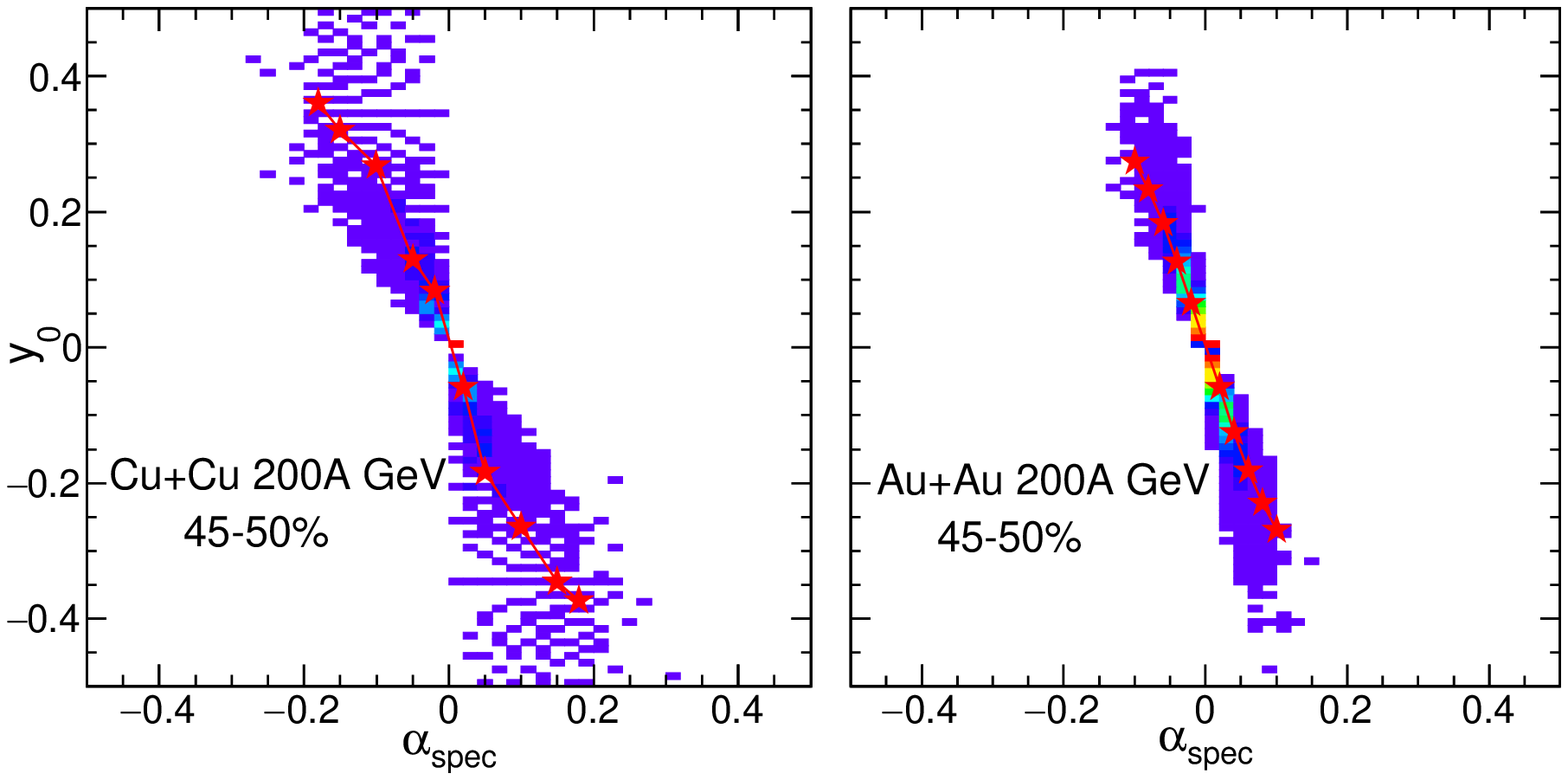}
	\caption{Event-by-event distribution of $y_{0}$ with $\alpha_{spec}$ from Cu+Cu and Au+Au collisions at 200A GeV for 0--5\% (top), 20--25\%(middle) and 45--50\% (bottom) centrality bins. Red line represents the mean $y_{0}$.}
\end{figure} 

Fig.2 shows the event-by-event distribution of $\alpha_{part}$ with $y_{0}$ from Cu+Cu and Au+Au collisions at two different centrality bins.  They are almost linearly correlated with positive slope. However, we do not see such strong linear correlation  in the distribution of $\alpha_{spec}$ and $y_{0}$ due to the presence of the (A+B) term in Eq. (5) (see Fig.3). There is a distribution of $y_0$ for a given value of $\alpha_{spec}$. The mean $\langle y_{0} \rangle$ is linearly correlated with $\alpha_{spec}$ with negative slope. The $\langle y_{0} \rangle$ does not change significantly for 0--5\% centrality bin over the whole range of $\alpha_{spec}$ for Au+Au collisions. The spread in $y_0$ increases for more peripheral collisions and also it is more for Cu+Cu collisions compared to Au+Au collisions.

It has been mentioned that the asymmetry in $\alpha_{spec}$ can be estimated by the spectators  energy deposition in the ZDCs. The variation of average rapidity shift with $\alpha_{spec}$ is shown in Fig.3 (red lines). The average rapidity shift has been found to be almost linearly related to the spectator asymmetry. The colours in Fig.2 and Fig.3 represent the event counts with red indicating most number of events and blue least number of events.

\section{Effect of rapidity shift on the rapidity distribution}
The rapidity distribution of charged particles produced in a collision of identical nuclei can be described by a symmetric Gaussian distribution about the center of mass rapidity. If the rapidity of the participant zone is shifted by $y_{0}$ from  the nucleon-nucleon center of mass system, then the modified  distribution in the participant zone is described by a  Gaussian whose mean is at $y_{0}$,
\begin{eqnarray}
\frac{dN}{dy} \ \propto \ e^{-\frac{(y-y_{0})^{2}}{2\sigma^{2}}}
\end{eqnarray}
\begin{figure}[h]
	\includegraphics[width=0.5\textwidth]{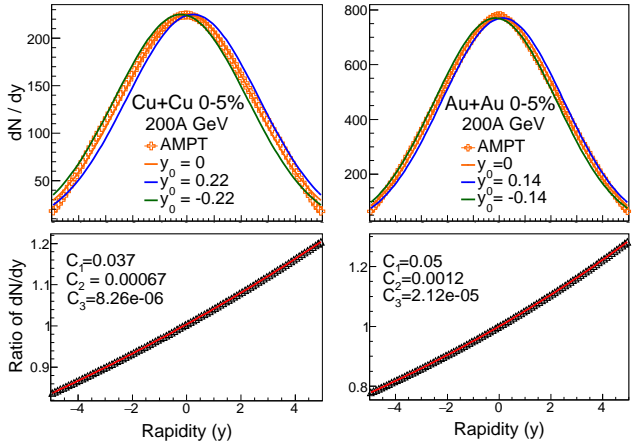}
	\caption{The $\frac{dN}{dy}$ distributions from Cu+Cu and Au+Au collisions for 0--5\% centrality bin [upper panel]. The ratio of the  distributions fitted to a third-order polynomial [lower panel]. }
\end{figure}
\begin{figure}[h!]
\includegraphics[width=0.5\textwidth]{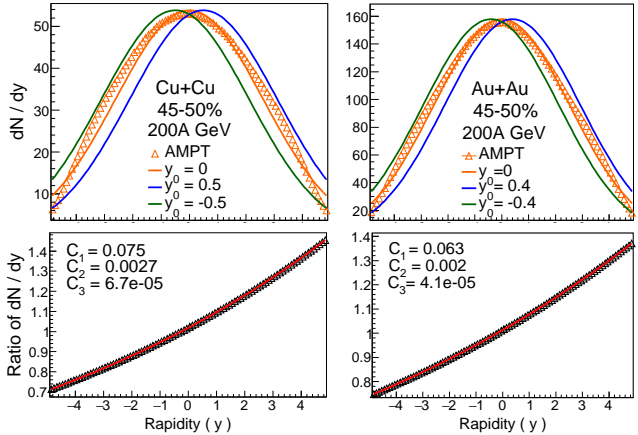}
\caption{The $\frac{dN}{dy}$ distributions  from Cu+Cu and Au+Au collisions for 45--50\% centrality bin [upper panel]. The ratio of the  distributions fitted to a third-order polynomial [lower panel].}
\end{figure}
where, $\sigma$ is the width of the Gaussian distribution. The positive and negative values of $y_{0}$ correspond to the positive and negative directions of net momentum of the participant zone respectively.

 Fig. 4 shows the shifted rapidity distributions with $y_{0}=\pm0.22$ and $y_{0}=\pm0.14$ for Cu+Cu 0--5\% and Au+Au 0--5\% collisions respectively. These $y_0$  values are the maximum shift in rapidity for the Au+Au and Cu+Cu collisions at 0--5\% centrality bin. The unshifted distributions  match with 
the data well~\cite{rap_dist}. 

One can see a marginal change in the rapidity distribution due to  $y_0$ shift for the most central collisions. In addition, the effect of rapidity shift is found to be slightly more for Cu+Cu collisions than for the Au+Au collisions at 200A GeV.

However, for peripheral collisions the $dN/dy$ distribution is significantly affected by the rapidity shift.  We show the shifted distribution for 45--50\% centrality bin in Fig. 5 for Au+Au and Cu+Cu collisions. The spread in the rapidity distribution is again found to be more for Cu+Cu collisions than for Au+Au collisions. These results show that longitudinal asymmetry is more pronounced for peripheral collisions and for smaller systems.

The ratio of the  unshifted to shifted distribution is shown in the bottom panels of the same figures. The ratio  is a valuable parameter to eliminate the effect of the uncertainties due to experimental corrections and fluctuations affecting event-by-event distribution. The ratio of the symmetric and asymmetric distributions can be expanded in a power series of $y$ as~\cite{alice, raniwala}
\begin{eqnarray}
\frac{dN/dy|_{sym}}{dN/dy|_{asym}}=\frac{f(y)}{f(y-y_{0})}\propto\sum_{0}^{\infty}c_{n}(y_{0},\sigma)y^{n}
\end{eqnarray}
where, $c_{n}$ are the coefficients of the Taylor expansion. One can see that the coefficients of the expansion depend on the shape  of the actual rapidity distribution and also on rapidity shift $y_{0}$.

The dominant term in the expansion is linear in $y$ which signifies the magnitude of asymmetry. The coefficients are extracted by fitting the ratio to a third-order polynomial~\cite{alice, raniwala}. The shift ($y_0$) increases for more glancing collisions which is manifested by  the coefficients at different centrality bins. Thus $c_1$ ($\propto  y_0$) is expected to be larger for larger asymmetry. It has been shown in earlier studies that $c_2$ ($\propto  y_0^2$) is two order of magnitude smaller than $c_1$ and $c_3$ is  even more smaller~\cite{raniwala}.

\begin{figure}[h!]
	\includegraphics[width=0.5\textwidth]{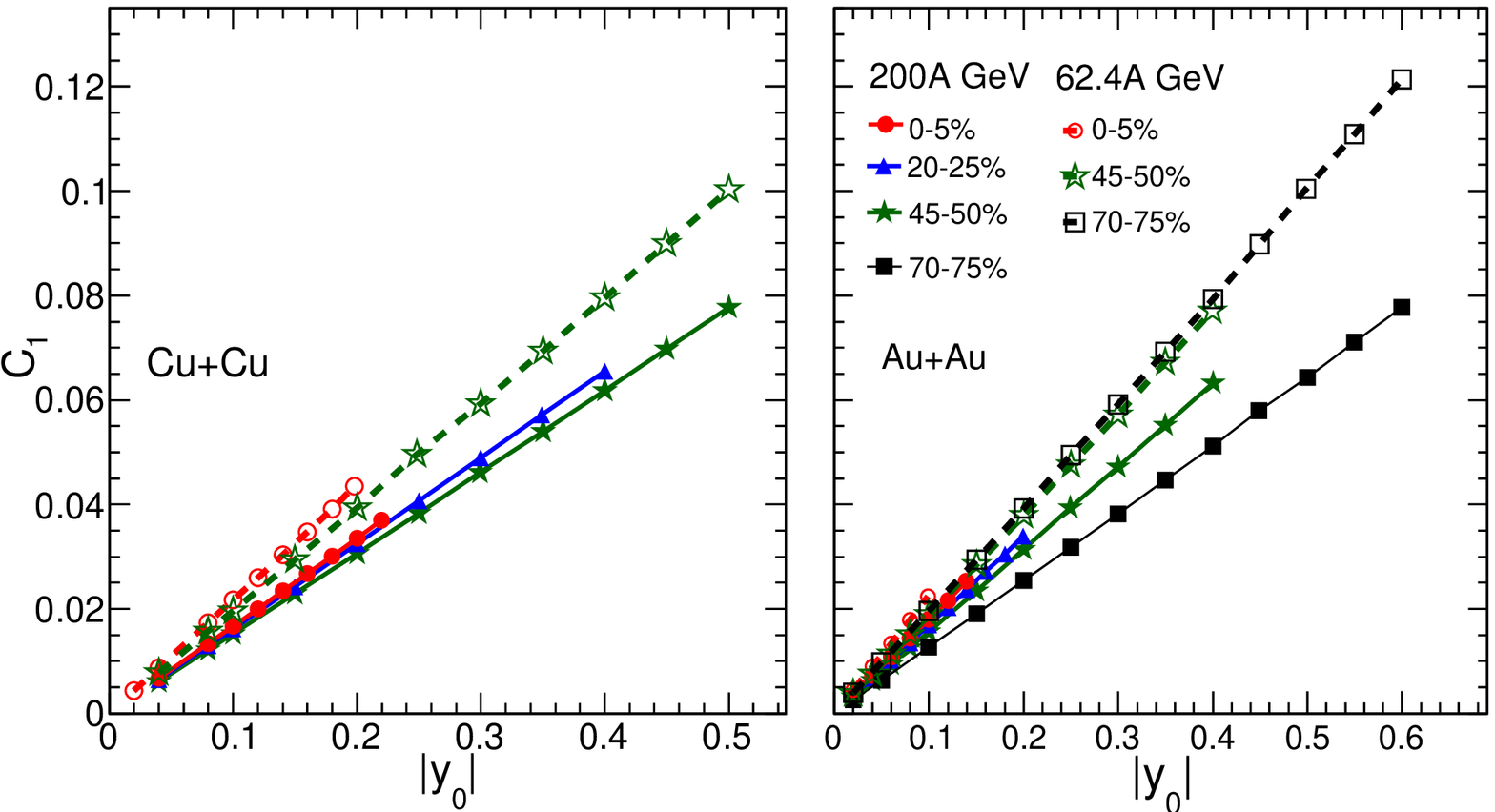}
	\hfill
%\end{figure}
%\begin{figure}[h]
	\includegraphics[width=0.5\textwidth]{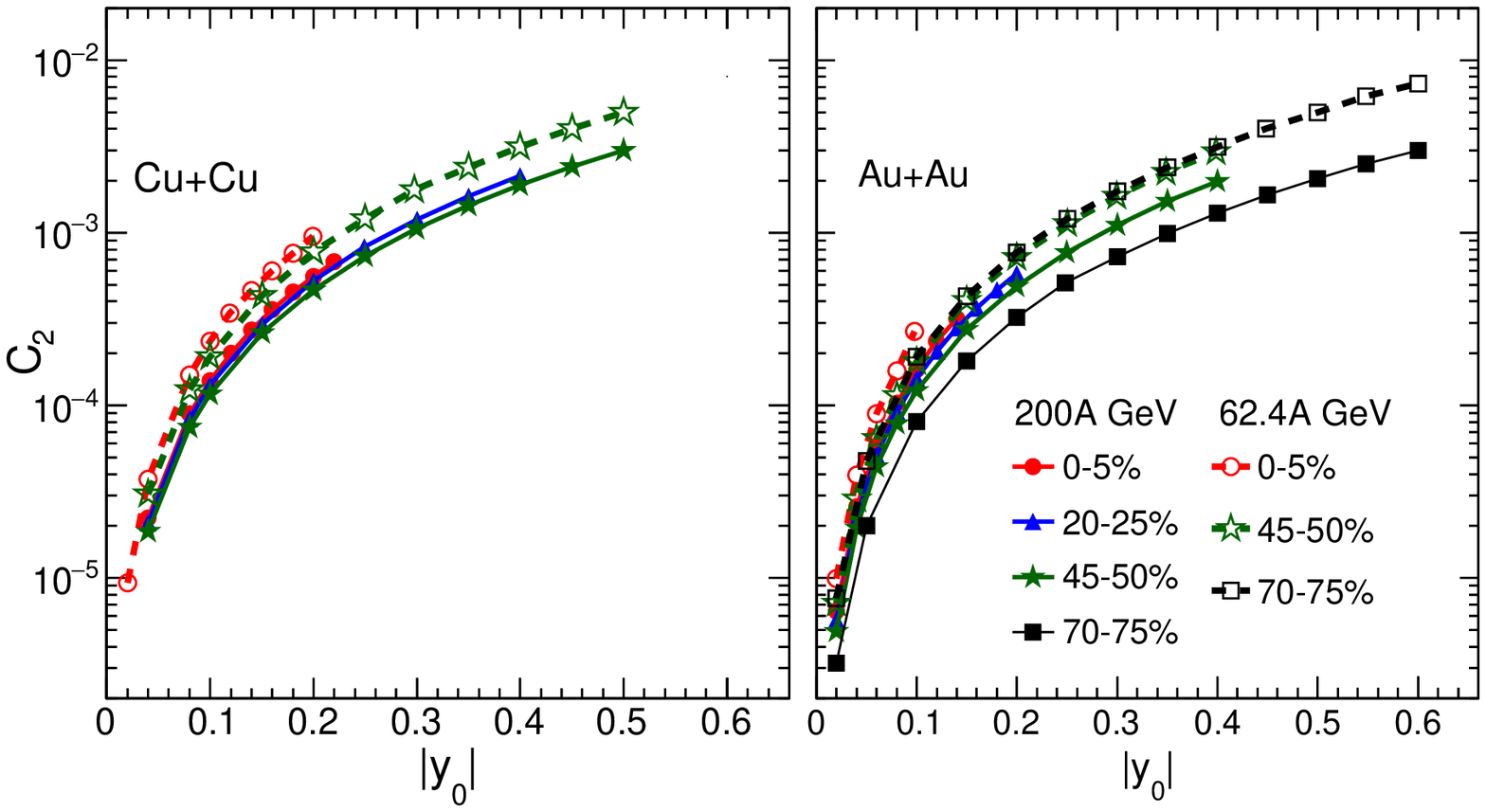}
	%\caption{text}
%\end{figure}
%\begin{figure}[h]
%	\includegraphics[width=0.5\textwidth]{Fig5c_paper.eps}
	\caption{The dependence of the coefficients $c_{1}$ and $c_{2}$  on rapidity shift $y_{0}$ from Cu+Cu and Au+Au at $\sqrt{s_{NN}}=$200 Gev and 62.4 GeV.}
\end{figure}
The dependence of the coefficients on the rapidity shifts is shown in Fig 6.  The results from Au+Au and Cu+Cu collisions at two different beam energies are shown for a comparison.
The coefficients vary significantly  for the two colliding systems at different centrality bins. The different values of $c_n$ indicate the sensitivity of the longitudinal asymmetry to the shape of rapidity distribution. 

The linear coefficient  is related to the rapidity shift as $c_{1}=\frac{y_{0}}{\sigma^{2}}$,  i.e. with increasing centrality, the width of the distribution decreases and as a result the slope of $c_{1}$ increases.
For 200A GeV Au+Au collisions, $c_1$ is found to be  larger for central collisions than for peripheral collisions. In addition, the same coefficient for 62.4 GeV Au+Au collisions is found to be larger compared to 200A GeV results. One can also see from the figures that  the spread in $y_0$ is slightly larger for lower beam energy for a particular centrality bin. 
For Cu+Cu collisions and Au+Au collisions we see a similar  qualitative nature of $c_1$. However,  a relative comparison shows that the $c_1$ values for Cu+Cu collisions are slightly smaller than the Au+Au results at  same centrality bin.

The coefficients for the quadratic and cubic terms  in the polynomial fitting of the ratio are found be significantly smaller than  the linear term, which  primarily determines the effect of asymmetry. However, the sensitivity of $c_2$ to the collision centrality and beam energy is clearly visible from the figures (see bottom panels of Fig. 6). %The $c_3$ coefficient also shows behaviour similar to $c_2$ (not shown here).

We have studied the energy dependence of the linear coefficient as a function of $\alpha_{\rm spec}$ and the results from  200A GeV are compared with 62.4 GeV for both  Cu+Cu and Au+Au collisions at RHIC (see Fig. 7). The results differ significantly when the beam energy is reduced to 62.4 GeV and it is more clearly visible when plotted with  $\alpha_{\rm spec}$ compared to $y_0$ (as shown in  Fig. 6).

The  coefficients show  stronger sensitivity to $\alpha_{spec}$ compared to the rapidity shift  and these results  show that the initial longitudinal asymmetry survives in the final state particle production.
\begin{figure}[h]
	\includegraphics[width=0.5\textwidth]{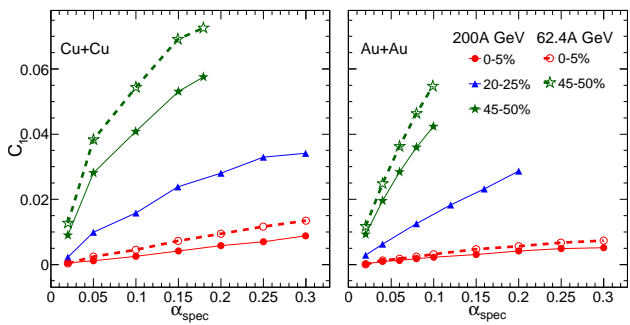}
	\caption{The dependence of the linear coefficient $c_{1}$ on spectator asymmetry $\alpha_{spec}$ for Cu+Cu and Au+Au at $\sqrt{s_{NN}}=$ 200 GeV and 62.4 GeV.}
\end{figure}
It is also clearly visible from Fig.7 that larger  asymmetry in the spectator nucleons results in larger value of the coefficient of the linear term. In addition, the $c_1$  strongly depends on the collision centrality. The asymmetry increases with decreasing centrality and also with decrease in the system size.

%The larger value of  the coefficient at 62.4 GeV  indicates an increase in fluctuation with decrease in the collision energy and as a result the difference increases for peripheral collisions.

\section{Effect of rapidity shift on flow coefficients}
The rapidity shift due to longitudinal asymmetry is also expected to affect the anisotropic flow parameters of the charged particle. We calculate the elliptic and the triangular flow parameters from each event by including the rapidity shift due to longitudinal asymmetry of the individual events. The final flow parameters are obtained by taking average over all the events in the centrality bin. The modified flow parameters are compared with the anisotropic flow parameters calculated without considering the rapidity shift (see Fig. 8). The results are shown for 45--50\% centrality bin where the effect of longitudinal asymmetry is expected to be more pronounced.

Both the elliptic and the triangular flow parameters are found be smaller when the rapidity shift is included. However, one can see from the figures that the effect of longitudinal asymmetry is clearly visible more for the triangular flow parameter compared to the elliptic flow parameter at RHIC. These results indicate that it would also be interesting to see the effect of rapidity shift on the directed flow parameter. However, we postpone that for a future study.

% As expected, we see a much smaller value of the flow parameters as a function of $p_T$  due to the shift in rapidity. The effect of the rapidity shift is seen to be more prominent for smaller systems and also for the triangular flow parameter compared to the elliptic flow of hadrons. In addition the $v_2$ and $v_3$ coefficients show larger variation for 62.4A GeV compared to 200 A GeV. Results from AMPT have been matched with data ~\cite{v2v3}.
\begin{figure}[h]
	\includegraphics[width=0.5\textwidth]{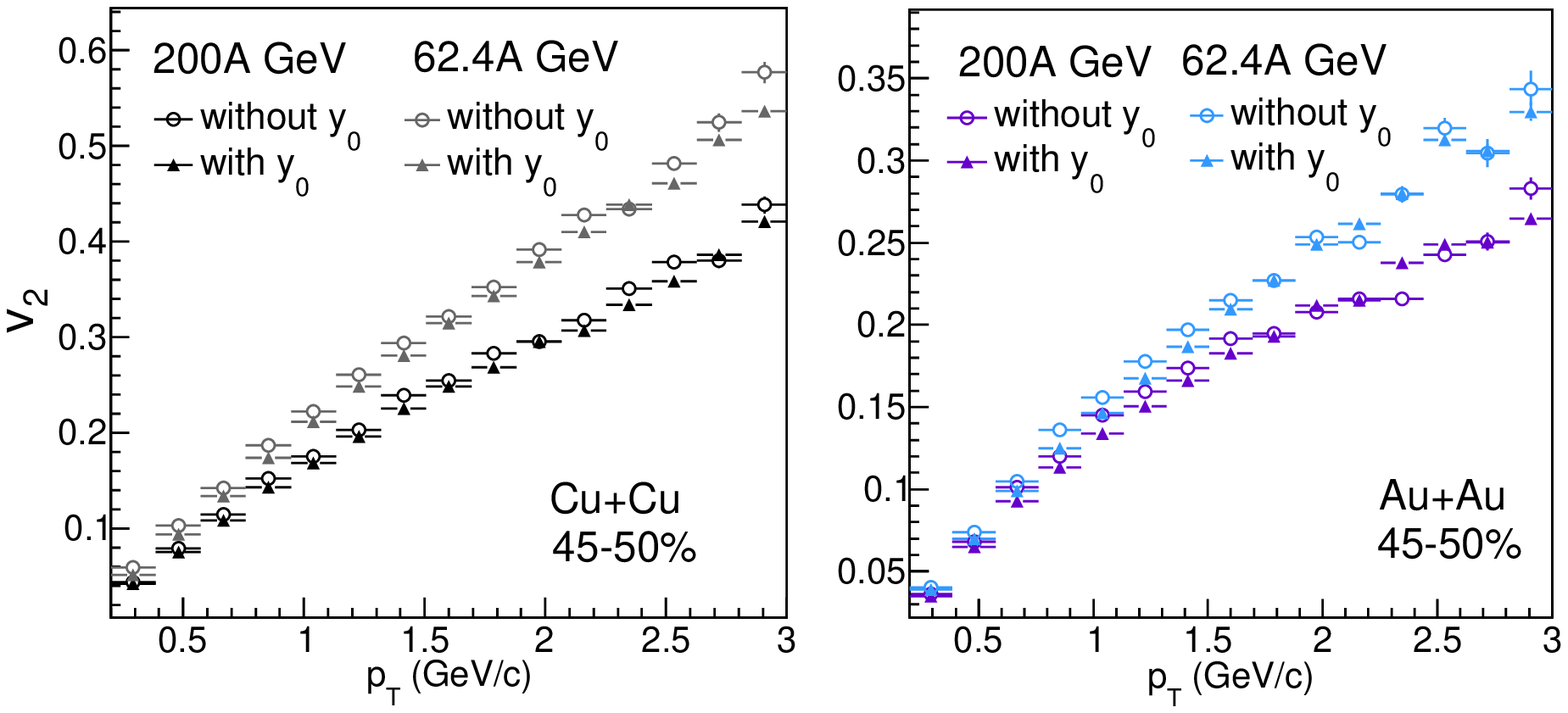}
	\vfill
	\includegraphics[width=0.5\textwidth]{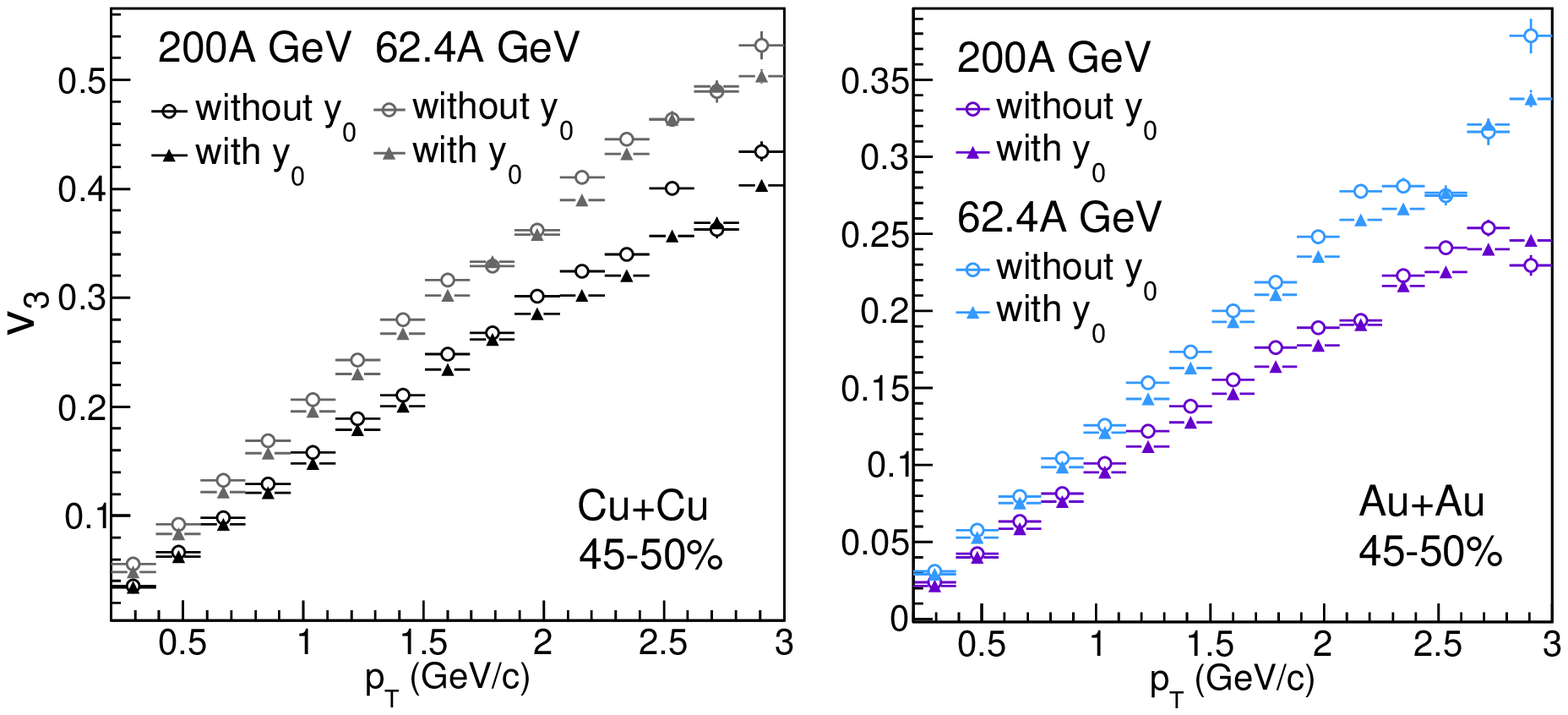}
	\caption{The effect of $y_{0}$ on the $p_{T}$ dependent flow coefficients $v_{2}$ and $v_{3}$ from Cu+Cu and Au+Au at $\sqrt{s_{NN}}=$200 GeV and 62.4 GeV at 45--50\% centrality bin.}
\end{figure}

\section{Summary and conclusions}
We study the effect of longitudinal asymmetry on hadronic observables from Au+Au and Cu+Cu collisions at RHIC energies for different centrality bins using the AMPT-SM model.  The rapidity shift of the participant zone due to longitudinal asymmetry is found to be quite sensitive to the system size, collision centrality and beam energy. The effect of longitudinal asymmetry is found to be more significant for peripheral collisions and for smaller systems.  The effect of the rapidity shift on the rapidity distributions of charged particles has been studied by taking the ratio of the $dN/dy$ of symmetric and asymmetries events. The ratio is described by a third-order polynomial. The large value of the coefficients of the polynomial confirms that  initial state longitudinal asymmetry can effect the final state observables. % The system size and beam energy dependence of the coefficients confirm that the longitudinal asymmetry is higher for smaller system and lower collision energies.
  The anisotropic flow parameters are also found to be affected by the longitudinal asymmetry. The triangular flow parameter is found to be affected more compared to the elliptic flow parameter. These results indicate that experimental analysis at RHIC energy would provide valuable information about the initial state and the effect of longitudinal asymmetry on the final state observables in heavy ion collisions.

\section{Acknowledgements}
We would like to thank the VECC-GRID computer facility and Abhisek Seal for their help in event generation and data storage. We thank Dr. Pingal Dasgupta for useful  discussion and help in Glauber model calculations.

\end{document}